\documentclass[conference]{IEEEtran}

\usepackage{cite}
\usepackage{amsmath,amssymb,amsfonts}
\usepackage{algorithmic}
\usepackage{graphicx}
\usepackage{textcomp}
\usepackage{listings}
\usepackage{xcolor}
\usepackage{subfig}
\usepackage{float} 

\def\BibTeX{{\rm B\kern-.05em{\sc i\kern-.025em b}\kern-.08em
    T\kern-.1667em\lower.7ex\hbox{E}\kern-.125emX}}
\begin{document}

\title{Reduction of Test Re-runs by Prioritizing Potential Order Dependent Flaky Tests\\}

\author{
\IEEEauthorblockN{Hasnain Iqbal}
\IEEEauthorblockA{University of Dhaka\\
Dhaka, Bangladesh \\
bsse1106@iit.du.ac.bd}
\and
\IEEEauthorblockN{Zerina Begum}
\IEEEauthorblockA{University of Dhaka\\
Dhaka, Bangladesh \\
zerinabegum@gmail.com}
\and
\IEEEauthorblockN{Kazi Sakib}
\IEEEauthorblockA{University of Dhaka\\
Dhaka, Bangladesh \\
sakib@iit.du.ac.bd}
}

\maketitle

\begin{abstract}
Flaky tests can make automated software testing unreliable due to their unpredictable behavior. These tests can pass or fail on the same code base on multiple runs. However, flaky tests often do not refer to any fault, even though they can cause the continuous integration (CI) pipeline to fail. A common type of flaky test is the order-dependent (OD) test. The outcome of an OD test depends on the order in which it is run with respect to other test cases. Several studies have explored the detection and repair of OD tests. However, their methods require re-runs of tests multiple times, that are not related to the order dependence. Hence, prioritizing potential OD tests is necessary to reduce the re-runs. In this paper, we propose a method to prioritize potential order-dependent tests. By analyzing shared static fields in test classes, we identify tests that are more likely to be order-dependent. In our experiment on 27 project modules, our method successfully prioritized all OD tests in 23 cases, reducing test executions by an average of 65.92\% and unnecessary re-runs by 72.19\%. These results demonstrate that our approach significantly improves the efficiency of OD test detection by lowering execution costs.
\end{abstract}


\section{Introduction}
The existence of flaky tests in test suites makes software testing unreliable. A flaky test generates inconsistent results running in the same environment and source code \cite{luo_first_fse}. Due to their non-deterministic nature, test suites with flaky tests can mislead developers into thinking of a flaw in a correctly executed code snippet\cite{parry_developers_icseseip}. Attempting to fix them frequently results in more bugs and slows down development. Developers confirmed that frequent encounters with flaky tests lead them to ignore potentially genuine failures\cite{parry_developers_icseseip}.

Luo et al.\cite{luo_first_fse} performed the first empirical study on flaky tests by studying bug reports in some open-source public projects. The study found that in Java projects, order-dependent (OD) flaky tests are among the top three most frequently observed types of flaky tests. OD flaky tests pass or fail based on the order in which they are executed relative to other tests\cite{lam_iDFlakies_icst,ishi_FixFlakies_fse,zhang_dtdetector_issta}. OD tests are deterministic tests concerning the test order, which either fail or pass in a particular order. And detecting these flaky tests (OD) is essential because they can mask true failures\cite{parry_survey_methodology}, reduce test reliability\cite{luo_first_fse}, and complicate debugging. Identifying these tests helps maintain the stability of test suite, ensures consistent results, and improves the effectiveness of automated testing\cite{li_incldflakies_issta}.


Detecting OD tests requires re-running the tests in multiple orders\cite{lam_iDFlakies_icst,ishi_FixFlakies_fse,li_base_issta,li_odrepair_icse,zhang_dtdetector_issta}. As a real-world project can contain numerous test cases, covering all possible orders to find an OD test can be impractical.

Several approaches are available for detecting and repairing Order Dependent (OD) flaky tests. Lam et al. previously developed \textit{iDFlakies}\cite{lam_iDFlakies_icst} to run tests on random orders \cite{lam_iDFlakies_icst}. Their technique was effective in detecting many OD tests. However, randomly shuffled test orders do not guarantee whether all OD tests are detected or not\cite{li_base_issta}. Order dependence can be present between any pair of tests in a suite. Therefore, the suite is to be run in all possible orders to detect all OD tests. Wei et al.\cite{wei_tuscan} later developed a more systematic way to execute test orders by covering pairs of tests. A test pair (t1, t2) is covered if there is a test order in which test t1 is positioned right before t2, with no other tests in between. This process minimizes the number of orders and detects all OD tests in the suite. Their approach is based on \textit{Tuscan Squares}\cite{etzion_tuscan}.

Later, Wei et al., Li et al.\cite{li_base_issta} proposed three different ways using the same \textit{Tuscan Squares} to minimize the test orders required to run. They created test orders with methods from the same class and cross-classes. They also targeted potential pairs for OD tests in an approach. However, there are still many tests to re-run that do not contribute to order-dependent flakiness. For example, according to their process, a test class with three test methods in it will need four orders for the test class to run. That makes each test method run four times, even if a test is not related to an OD test. This number will increase in real-world projects having more test cases. These redundant re-runs are computationally costly and have no use case other than the detection of OD tests. Therefore, prioritization of potential OD tests and reduction of unnecessary redundant test cases are required.

In this paper, we present an approach to prioritize potential order-dependent (OD) tests in Java projects, aiming to minimize the number of test re-runs required for detecting OD tests. Our focus is specifically on shared mutable states within test classes, accessible through static fields, which is recognized as the primary cause of order-dependent flakiness \cite{parry_survey_methodology, lam_iDFlakies_icst}. Gambi et al. \cite{gambi_practical}, in their work on PRADET, used dynamic data-flow analysis to uncover conflicting reads and writes on objects in the heap, systematically navigating through shared static fields to identify dependencies. Based on these insights, our static-analysis based method prioritizes potential OD tests by taking advantage of the dependencies discovered in prior studies \cite{gambi_practical, zhang_test-independence}. The effectiveness of our approach is evident, as it successfully prioritizes all OD tests in 23 out of 27 projects, achieving an average test reduction of 65.92\% per project.

\section{Background}
In this section we look into the order dependent flaky tests. We provide definitions and terms related to our work.

\subsection{Order-Dependent Flaky Tests:}
A test order refers to the order of the test methods in a test suite. An order-dependent (OD) test is a flaky test whose result (pass or fail) can change based on the order it runs with other tests. This means that shuffling the test methods in test order can make it pass or fail.
For an OD test, there is at least one order in which the test will pass (called a passing test order) and one order in which it will fail (a failing test order)\cite{li_base_issta}. A test to be an OD test must consistently produce the same result if run in the same order. If the result of a test case changes for reasons other than order, such as timing issues or external data, it is still considered flaky\cite{luo_first_fse,parry_survey_methodology}, but not order-dependent. Therefore,  OD tests are deterministic if the order is fixed\cite{lam_iDFlakies_icst,ishi_FixFlakies_fse,luo_first_fse}.

Shi et al.\cite{ishi_FixFlakies_fse} formalized categories of order-dependent flaky tests. They classified the OD tests into two types, \textit{victims} and \textit{brittles}. They also classified some tests related to OD as \textit{polluters}, \textit{cleaners}, and \textit{state setters}.

The most frequent type of OD tests are victims\cite{parry_survey_methodology}. Tests that fail while running in the order but pass when run in isolation (not within any order or any other tests) are called victims. A victim fails in the order if another test is running before it and modifying some shared states needed for the victim to pass. These shared states are shared static fields in Java\cite{lam_iDFlakies_icst}. These problematic state-modifier tests are called polluters. The polluter modifies or \textit{pollutes} some shared state between the polluter and the victim. And the victim fails for not having the expected state. However, there are some tests found that clear or reset the modified states by the polluter. When these tests are run between polluters and victims, they reset the problematic shared state, and the victims pass successfully. These tests are called cleaners. A cleaner cleans the shared states between a polluter and a victim. A victim can have multiple polluters and multiple cleaners.

Brittles are not as common as victims \cite{ishi_FixFlakies_fse} but still cover a significant portion of OD tests. To pass, a brittle needs another test to run before it. When a brittle runs in isolation, it fails, but when run after a particular test, it passes. The needed previous test is called a state setter. A state setter sets a value to the shared state that would be required for the brittle to pass. There can be multiple state setters for a brittle. A passing test order must have at least one state setter run before the brittle, and a failing test order would have no state setter run before the brittle.

\subsection{OD Test Detection and Fixing}
Several works have been done on detecting OD tests. Previous works proposed re-running tests in random orders\cite{lam_iDFlakies_icst,li_incldflakies_issta}. An OD test will have a passing test order as well as a failing test order. As the method implies, there is no guarantee of finding the both orders in a limited number of re-runs\cite{li_base_issta}. Without finding both passing and failing orders we can not confirm a test to be an OD or not.
Often, the victim contains multiple cleaners in a test suite. If the number of cleaners is much greater than polluters, the probability of finding a failing order for a victim becomes very low\cite{li_base_issta}. Because in maximum orders a cleaner would run after the polluter. Therefore, the suite needs to be run on many test orders to detect a victim. For some projects, the chance of finding a failing order can only be 1.2\%\cite{ishi_FixFlakies_fse}. It can similarly be difficult to detect a brittle as well. Therefore, if we consider the worst case to detect an order-dependent test, the test suite is needed to run on all permutations of the test cases, which is often not practical for a large project\cite{lam_iDFlakies_icst,li_base_issta}. For example, in our work, an open source project, jackson-databind, has 3624 tests. Running the test suites of this project in all possible permutations will be a costly approach to detect OD tests.

To solve this problem, Wei et al.\cite{wei_tuscan} used the Tuscan Square\cite{etzion_tuscan} theory to generate test orders. It guarantees that all test pairs are covered in \textit{N} or \textit{(N + 1) }test orders for any \textit{N} tests. If every pair of tests is run consecutively, both victim and brittle cases get checked. If there is any victim, a polluter must have run before it in some order without the interference of a cleaner. And for the brittle, there definitely will be two cases where a state setter runs before it and does not run before it. Wei et al. used Tuscan Square to shuffle the methods of a class for generating orders.

Li et al.\cite{li_base_issta} proposed a reliable approach for detecting OD tests with both computational efficiency and precision. They used the Tuscan Square theory to propose three different methods: \textit{Tuscan Intra-Class}, \textit{Tuscan Inter-Class}, and \textit{Target Pairs}. Among these, \textit{Tuscan Intra-Class }demonstrated the highest effectiveness, identifying 97.2\% of OD tests in their study. In this method, they used the Tuscan Square to shuffle the methods within a class, alongside shuffling the classes. They increased the accuracy to 97.2\% from 36\% of Wei et al.\cite{wei_tuscan}. The required number of orders also does not vary much. On average, 94.0 orders were generated by the method of Wei et al. whereas 104.7 orders were generated for the Intra-Class method.

{Need of Test Method Prioritization}
\textit{Tuscan Intra-Class} employs a minimal number of orders to detect order-dependent (OD) tests. However, each test order requires re-running all the test cases within a class, even though many of these tests do not contribute to order dependence (i.e., they are not victims, brittle tests, polluters, cleaners, or state setters). 

Figure 1 shows a test suite with four test cases, where one (B) is an OD-victim. This setup results in eight unnecessary re-runs in Tuscan orders. In the figure, A, B, C, and D represent test cases. A is a polluter, B is the OD-victim, and C and D are non-contributing tests. Re-running C and D does not help detect the OD test. Only A and B need to be run multiple times to identify it. In this example, an OD test ratio of 25\% (1 in 4) results in 50\% (8 out of 16) of the test runs being unnecessary. In real-world projects, where OD tests make up much less portion of total test cases \cite{luo_first_fse,lam_iDFlakies_icst}, the proportion of unnecessary re-runs would be even higher, as reflected in our results. Keeping only the potential OD tests in the orders can minimize the unnecessary re-runs of non-contributing tests without compromising the detection of OD tests. However, no such work has yet been done.

\begin{figure}[htbp]
    \centering
    \subfloat[Run in Tuscan Order]{%
        \includegraphics[width=0.23\textwidth]{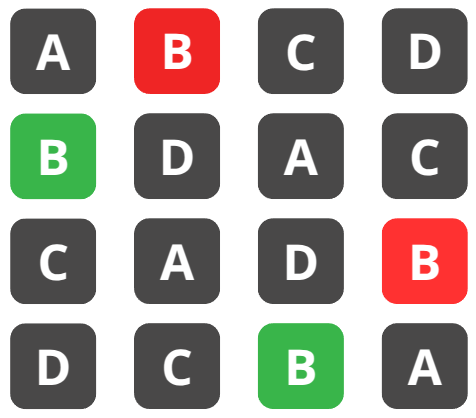}%
    }
    \hfill
    \subfloat[Unnecessary Re-runs]{%
        \includegraphics[width=0.23\textwidth]{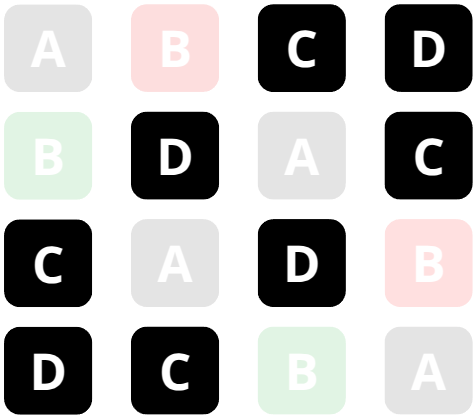}%
    }
    \caption{Volume of unnecessary re-runs}
    \label{fig}
\end{figure}

Detecting contributing tests can be challenging. As OD tests are triggered by shared in-memory heap states (in Java projects), a test might modify a shared state somewhere in its execution, creating potential order dependencies\cite{parry_survey_methodology}. Therefore, we need to analyze a test’s execution trace to confirm whether it accesses shared memory. In this study, we try to predict contributing tests using static analysis only on the test classes. Our results are promising, demonstrating the feasibility of test class based approach. These contributing tests are prioritized to generate orders.

When two or more code snippets rely on a shared memory state, the potential for order dependence arises. We observed that in nearly every instance, OD test pairs (such as victim-polluter or brittle-state setter) shared one or more static attributes from the test class. Although some non-OD tests also had shared states, these tests were likely candidates for OD pairs. This observation formed the foundation of our prioritization method.

In Figure 2, we can see an example test class from project Activity. At first, a static variable is declared \textit{currentTaskId} in line 4. This static variable is shared between both \textit{bCreateStandaloneTask} (line 14) and \textit{ctryCompletingWithUnauthorizedUser} (line 23) tests.
We can not confirm before executing the test orders, but it is clear that these two functions are potential OD candidates. Our method detects all shared static fields with the related test methods. These methods are marked as potential OD candidates, and they are prioritized over test methods without having shared static fields. Based on this prioritized list, the system generates test orders from a test class to detect OD tests. This process reduces unnecessary tests from the orders.
\begin{figure}[H]
\vspace{5pt}
\begin{lstlisting}
  public class TaskRuntimeCompleteTaskTest {
  
      private static String currentTaskId;
      ...
  
      @Test
      ...
      public void bCreateStandaloneTask() {
          ...
          assertThat(tasks.getContent());
          Task task = tasks.getContent().get(0);
          ...
          currentTaskId = task.getId();
      }
      ...
      @Test(expected = NotFoundException.class)
      ...
      public void ctryCompletingWithUnauthorizedUser() {
          taskRuntime
          .complete(TaskPayloadBuilder
            .complete()
            .withTaskId(currentTaskId)
            .build()
          );
      }
      
      ...
  }
\end{lstlisting}
\centering
\caption{Probable Order Dependent Test Pair, Test Class from project Activity}
\end{figure}
\section{Methodology}
In our approach, we prioritize test cases based on the states they require. We explore the shared static variables between tests.

\subsection{AST Generation}
The Abstract Syntax Tree (AST) is a powerful tool in programming and computer science to represent and analyze code structure in a hierarchical tree-like format. By breaking down code into its syntactical components, ASTs allow efficient manipulation, analysis, and transformation of code without executing it. Generating an Abstract Syntax Tree (AST) for a project is an effective way to analyze its structure. For the projects in our experiment, we used a well-regarded Java AST parser, ASTExtractor\cite{astextractor}, for generating AST. We derived ASTs for all related files for further analysis.

\subsection{Prioritizing Pairs}
For each test method, we determine the static variables accessed by the methods and keep it in a key-value pair. If any method within the class shares the same static field we mark both the methods as prioritized. We can see the pseudo code on Figure 3.
\begin{figure}[H]
\vspace{5pt}
\centering
\begin{lstlisting}
  ast = generate_ast(module)
  for each test file:
    for all fields in ast:
        if field is static:
            methods = find_methods(field)
        for each method:
            shareds[method] = {field, file}

  for method1 and method2:
    if c in shared[method1] and shared[method2]:
        priority_list.append(method1, method2)
\end{lstlisting}
\caption{Pseudo Code of Prioritizing Candidate OD Tests}
\end{figure}

\subsection{Test Order Generation}
Running in various test orders is required to detect the order-dependent flaky tests. Generating the minimum number of orders to detect all OD tests requires a Tuscan Square\cite{li_base_issta} approach. This method structures test classes in such an order that all test class pairs are covered at least once. Without the Tuscan Square approach, covering all test orders would require testing all permutations of test methods (and test classes). By considering the prioritized test cases with the relevant test classes, we generate intra-class orders suggested in \cite{li_base_issta}. Using the prioritized tests lowers the length of test orders and significantly reduces test re-runs.

\section{Experimental Setup}

We used data from the widely used IDoFT\cite{InternationalDatasetofFlakyTests,lam_iDFlakies_icst} dataset in our experiment. From IDoFT, we only took OD tests with accepted pull requests from developers as OD tests. We evaluated 27 project modules having 189 confirmed OD tests.
Our 27 selected modules had a total of 2545 test classes with an average of 86.82 classes per one. We had 13909 test methods, averaging 502.36 test methods per module and 8.21 tests per class. Among these, 189 were marked OD tests, which is 1.35\% of the total test cases. We can see a summary of the project modules used in our experiment in Table I.
\begin{table}[ht]
\renewcommand{\arraystretch}{1.5} 
\centering
\caption{Subjects of the Experiment}
\begin{tabular}{|c|p{3.9cm}|c|c|c|c|}
\hline
\textbf{ID} & \textbf{Module} & \textbf{\#Classes} & \textbf{\#Tests} & \textbf{\#OD} \\
\hline
1 & admiral-compute & 91 & 926 & 1 \\ \hline
2 & admiral-request & 54 & 563 & 1 \\ \hline
3 & aismessages & 19 & 49 & 2 \\ \hline
4 & biojava-structure & 100 & 492 & 1 \\ \hline
5 & dubbo-1f84cdc & 280 & 1367 & 10 \\ \hline
6 & dubbo-common & 159 & 811 & 1 \\ \hline
7 & dubbo-config-api & 64 & 395 & 37 \\ \hline
8 & easyexcel & 156 & 500 & 2 \\ \hline
9 & hadoop-hdfs-nfs & 14 & 96 & 28 \\ \hline
10 & hadoop-mapreduce-client-app & 59 & 739 & 4 \\ \hline
11 & hadoop-mapreduce-client-core & 61 & 462 & 5 \\ \hline
12 & hadoop-mapreduce-client-hs & 30 & 315 & 2 \\ \hline
13 & hdfs-nfs-a585a7 & 14 & 96 & 1 \\ \hline
14 & hdfs-nfs-cc2babc & 14 & 96 & 20 \\ \hline
15 & hippo4j-common & 40 & 196 & 1 \\ \hline
16 & http-request-lib & 3 & 168 & 25 \\ \hline
17 & incubator-ratis-server & 40 & 428 & 2 \\ \hline
18 & jackson-databind & 612 & 3524 & 1 \\ \hline
19 & jboot & 112 & 250 & 6 \\ \hline
20 & jnr-posix & 26 & 183 & 6 \\ \hline
21 & light-4j-correlation & 1 & 6 & 1 \\ \hline
22 & activiti & 13 & 53 & 16 \\ \hline
23 & secor & 35 & 172 & 2 \\ \hline
24 & spring-boot-actuator-autoconfigure & 180 & 905 & 8 \\ \hline
25 & spring-boot-test-autoconfigure & 286 & 592 & 4 \\ \hline
26 & undertow-jsr & 49 & 158 & 1 \\ \hline
27 & unix4j-command & 33 & 367 & 1 \\ \hline
\textbf{Sum} & \textbf{Total} & \textbf{2545} & \textbf{13909} & \textbf{189} \\ \hline
\end{tabular}
\label{table:data-overview}
\end{table}

\section{Evaluation}
\begin{table*}[ht]
\renewcommand{\arraystretch}{1.5} 
\centering
\caption{Reduction and Prioritization Analysis on Subjects}
\resizebox{\textwidth}{!}{ 
\begin{tabular}{|c|c|c|c|c|c|c|c|c|c|}
\hline
 & \multicolumn{3}{c|}{\textbf{Tuscan Intra-Class Order}} & \multicolumn{3}{c|}{\textbf{Prioritized Intra-Class Order}} & \multicolumn{3}{c|}{\textbf{Prioritization Result}} \\
\hline
\textbf{ID} & \textbf{\#Tests} & \textbf{Avg M/C} & \textbf{\#Test Run Needed} & \textbf{\#Tests} & \textbf{Avg M/C} & \textbf{\#Test Run Needed} & \textbf{OD Covered (\%)} & \textbf{Test Reduced (\%)} & \textbf{Test Run Reduced (\%)} \\
\hline
1 & 926 & 10.18 & 9422.81 & 424 & 4.66 & 1975.56 & 100 & 54.21 & 79.03 \\ \hline
2 & 563 & 10.43 & 5869.8 & 171 & 3.17 & 541.5 & 100 & 69.63 & 90.77 \\ \hline
3 & 49 & 2.58 & 126.37 & 7 & 0.37 & 2.58 & 100 & 85.71 & 97.96 \\ \hline
4 & 492 & 4.92 & 2420.64 & 131 & 1.31 & 171.61 & 100 & 73.37 & 92.91 \\ \hline
5 & 1367 & 4.88 & 6673.89 & 270 & 0.96 & 260.36 & 90 & 80.25 & 96.1 \\ \hline
6 & 811 & 5.1 & 4136.61 & 143 & 0.9 & 128.61 & 100 & 82.37 & 96.89 \\ \hline
7 & 395 & 6.17 & 2437.89 & 98 & 1.53 & 150.06 & 100 & 75.19 & 93.84 \\ \hline
8 & 500 & 3.21 & 1602.56 & 414 & 2.65 & 1098.69 & 100 & 17.2 & 31.44 \\ \hline
9 & 96 & 6.86 & 658.29 & 52 & 3.71 & 193.14 & 89.29 & 45.83 & 70.66 \\ \hline
10 & 739 & 12.53 & 9256.29 & 549 & 9.31 & 5108.49 & 100 & 25.71 & 44.81 \\ \hline
11 & 462 & 7.57 & 3499.08 & 286 & 4.69 & 1340.92 & 100 & 38.1 & 61.68 \\ \hline
12 & 315 & 10.5 & 3307.5 & 248 & 8.27 & 2050.13 & 100 & 21.27 & 38.02 \\ \hline
13 & 96 & 6.86 & 658.29 & 52 & 3.71 & 193.14 & 100 & 45.83 & 70.66 \\ \hline
14 & 96 & 6.86 & 658.29 & 52 & 3.71 & 193.14 & 100 & 45.83 & 70.66 \\ \hline
15 & 196 & 4.9 & 960.4 & 38 & 0.95 & 36.1 & 100 & 80.61 & 96.24 \\ \hline
16 & 168 & 56 & 9408 & 166 & 55.33 & 9185.33 & 100 & 1.19 & 2.37 \\ \hline
17 & 428 & 10.7 & 4579.6 & 296 & 7.4 & 2190.4 & 100 & 30.84 & 52.17 \\ \hline
18 & 3524 & 5.76 & 20291.79 & 647 & 1.06 & 684 & 100 & 81.64 & 96.63 \\ \hline
19 & 250 & 2.23 & 558.04 & 13 & 0.12 & 1.51 & 66.67 & 94.8 & 99.73 \\ \hline
20 & 183 & 7.04 & 1288.04 & 139 & 5.35 & 743.12 & 100 & 24.04 & 42.31 \\ \hline
21 & 6 & 6 & 36 & 6 & 6 & 36 & 100 & 0 & 0 \\ \hline

23 & 172 & 4.91 & 845.26 & 43 & 1.23 & 52.83 & 100 & 75 & 93.75 \\ \hline
24 & 905 & 5.03 & 4550.14 & 109 & 0.61 & 66.01 & 100 & 87.96 & 98.55 \\ \hline
25 & 592 & 2.07 & 1225.4 & 83 & 0.29 & 24.09 & 100 & 85.98 & 98.03 \\ \hline
26 & 158 & 3.22 & 509.47 & 95 & 1.94 & 184.18 & 100 & 39.87 & 63.85 \\ \hline
27 & 367 & 11.12 & 4081.48 & 173 & 5.24 & 906.94 & 100 & 52.86 & 77.78 \\ \hline
\textbf{Sum/Avg} & \textbf13909 & 8.21 & 99277.99 & 4740 & 5.08 & 27612.68 & \textbf{96.61} & \textbf{65.92} & \textbf{72.19} \\ \hline
\end{tabular}
}
\label{table:reduction-analysis}
\end{table*}
Our evaluation addresses the following research questions:

\textbf{RQ1:} How effective is our approach in reducing test case re-runs?

\textbf{RQ2:} In terms of prioritization accuracy, is analyzing test classes enough to prioritize tests to detect OD tests?

\subsection*{RQ1: Reducing test case re-runs}
In the Tuscan Intra-Class method, both the classes and their test cases are shuffled to make orders. However, it can be done parallelly. Here, we present a mathematical analysis of the number of re-runs of the Tuscan Intra-Class method. We assume that each order will consist of one class only. Then the needed number of orders (\textit{TIO}) in a project will be,
\begin{equation}
TIO = max(T(C), {T(M_c)})
\end{equation}
Where $C$ stands for the number of Classes, and $M_c$ stands for the number of Methods concerning its class. $T(C)$ represents the number of Tuscan orders needed for classes, and $T(M_c)$ represents the total number of Tuscan orders needed for methods within classes. Since the number of methods will always be greater than or equal to the number of classes, we can rewrite it as follows:
\begin{equation}
TIO = T(M_c)
\end{equation}
Now,
\begin{equation}
T(M_c) = \sum_{i=1}^{C} T(M_i)
\end{equation}
The average number of methods for each class as $M/C$, then it simplifies to,
\begin{equation}
TIO = C*T(M/C)
\end{equation}
We know that, for a natural number N, Tuscan Square length is N or N+1, For simplicity, we take N for all and the equation becomes,
\begin{equation}
TIO \simeq C*(M/C) = M
\end{equation}
Therefore, the number of test run (re-run) in a project according to Tuscan Intra-Class method will be,
\begin{equation}
N \simeq TIO*M/O
\end{equation}
Methods per order (M/O) will be the method count of the respective order. As we assumed the order size is the same as the class size (number of methods), we can simplify it to the average number of methods per class.
\begin{equation}
N \simeq TIO*M/C = M*M/C = M^2/C
\end{equation}
Using this equation, we calculate the number of required runs for tests that are shown in Table II.

As discussed earlier, \textit{Tuscan Square} based Intra-Class OD detection is the most reliable and computationally friendly OD detection technique. It can detect 97.2\% OD test successfully with the least number of test cases that need to be re-run. However, it still needs to run a lot (as we discussed earlier) of extra test cases that do not affect detecting OD. Using our method of reducing the number of test cases in orders can decrease the number of test cases to re-run without hampering the detection reliability. Our shortlist of test cases consists of all possible OD tests. They can be detected by any other detection mechanism. Our experiment shows that we can reduce 65.92\% of test cases from all over the projects. That decreases the number of required re-runs by 72.19\% from all the projects. 

\begin{figure*}[htbp]
\centerline{\includegraphics[width=\textwidth]{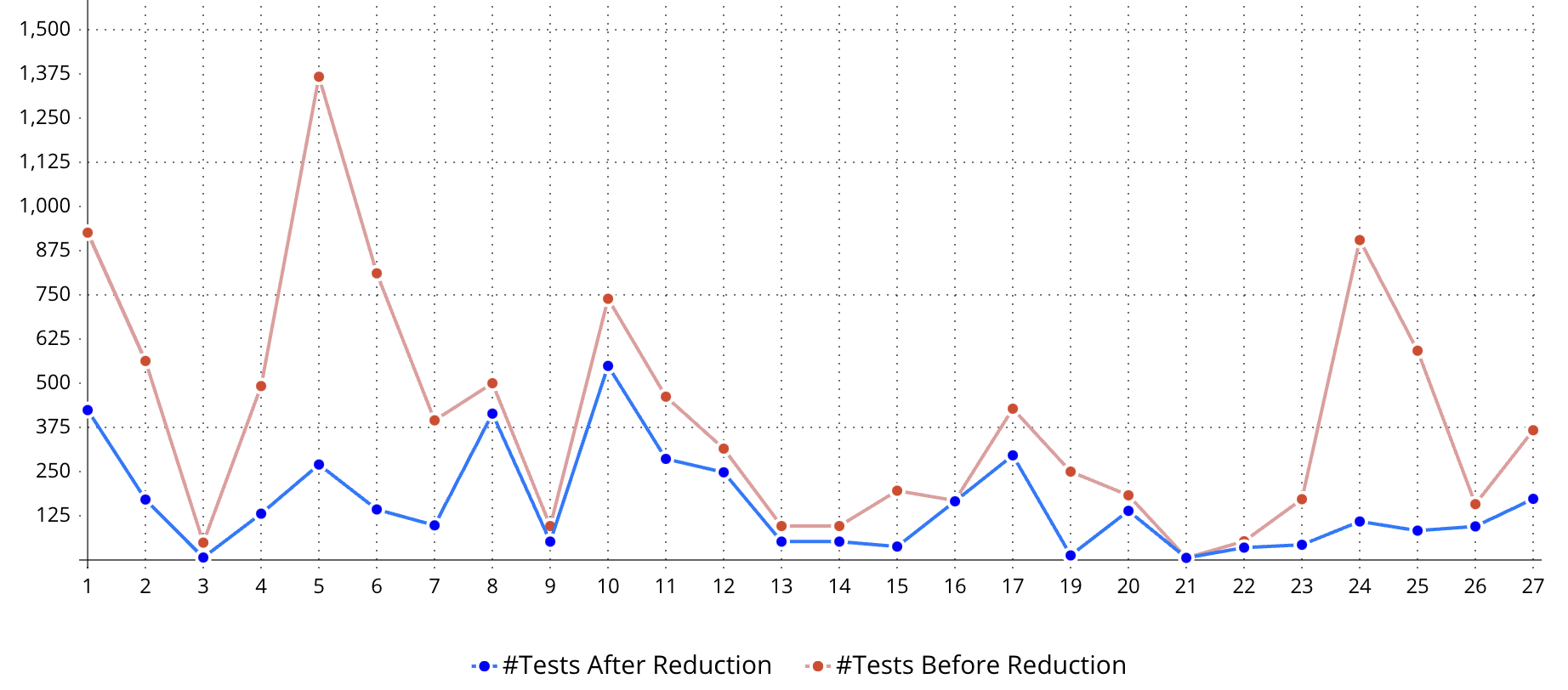}}
\caption{Number of test cases before and after reduction}
\label{fig}
\end{figure*}

\subsection*{RQ2: Test classes in the accurate prioritization of potential OD tests}
As we can see in our result, only analyzing test classes can be a prolific option in terms of efficiency. Multiple Recent studies suggest that issues in test classes have a significant effect on OD tests. Parry et al. found setup and teardown to be the most common causes of flaky tests\cite{parry_developers_icseseip}. Shi et al. used test methods to generate patches for OD tests in iFixFlakies\cite{ishi_FixFlakies_fse}. They found patches for 58 out of 110 OD tests.
In our study of 27 project modules, 23 of them had successful prioritization of all 100\% OD tests in the module. On average, 65.92\% of the total test cases reduced. However, four modules show that we have lost some OD. Although we got 177 out of 189 OD tests, which is 96.61\% accurate, for reliability we always want this to be 100\%. This matter raises a concern about how effective it will be in real-world industry projects. 
Our analysis on this issue found that these modules have lost some OD tests because of the code snippet under test modifies another shared state. This suggests that we can prioritize more accurately if we consider the code under test also, and we are currently working on this to reach a conclusion. However, the test classes show much potential in prioritizing OD tests by securing complete reliability on 23 out of 27 project modules.

\section{Threats to Validity}
We got our results from a small study of 27 project modules. Thus, it may not be generalized to other subjects. Also, we depended on the data from previous works. As the data are rechecked and multiple prior successful works have been done, an issue with data is unlikely to happen. According to our check using the method from the previous study\cite{li_base_issta}, we find it as the same as the dataset shows.

Also, we have removed non-contributing test cases from the order. As our study is designed for detecting order-dependent flaky tests, removing non-contributing tests does not hamper the testing process. To ensure reliability, we must run the whole test suit after mitigating the order dependence issue.

Finally, as we focused only on the test classes, projects with issue in the code under test will be untouched.

\section{Related Works}
Flaky tests have been explored widely for the past 10-12 years. Different mechanisms of detection and fixing have been proposed to mitigate test flakiness.
Luo et al.\cite{luo_first_fse} with one of the first empirical studies of flaky tests in open source projects. They worked on 201 commits of 51 open-source projects to classify the most common root causes. Their findings on order-dependent tests being in the top three causes of test flakiness put it to the light for the researchers to work on.
Lam et al. proposed iDFlakies\cite{lam_iDFlakies_icst} for detecting OD and non-OD tests. In their study, they found which tests deterministically fail in any random order different from the original passing order. If any test passes in the original order but fails deterministically in another order, it is classified as an OD test. If it non-deterministically passes or fails in an order it's classified as a non-OD test. This method is effective in finding OD tests. However, finding the exact order in which it fails may require an unfixed number of retries. Thus, it cannot provide the guarantee to find all of the OD tests.
Wei et al.\cite{wei_tuscan} later provided a solution to this problem without running all possible orders, which would be computationally costly. They used the Tuscan Square approach to cover all possible test pairs in N or N+1 number of test orders for any N tests. This method guarantees the execution of each test case of a class to be tested either OD or not. However, their method is limited to the tests within a test class. It cannot detect the OD tests which are caused by a method from another class.
Li et al.\cite{li_base_issta} used this Tuscan square approach to run between classes as well resulting in a much more reliable way to detect OD tests. The run Tuscan order in between classes also covers all possible dependencies of a test case. They proposed three ways of detecting, \textit{Tuscan Intra-Class}, \textit{Inter-Class}, and \textit{Target Pairs}. Among these, \textit{Tuscan Intra-Class} reportedly performs the best. It can detect 97.2\% of OD tests with the number of test orders needed similar to \cite{wei_tuscan}.
Static shared state based detection works have been widely explored previously. Gambi et al. \cite{gambi_practical} and Zhang et al. \cite{zhang_test-independence} provided the base of using shared static fields and our idea of prioritizing OD tests is based upon their contribution.
Some other studies also explored the detection of OD tests. DTDetector\cite{zhang_dtdetector_issta}, proposed by Zhang et al., detects OD tests by re-running tests in random test orders or running pairs of tests. IncIDFlakies\cite{li_incldflakies_issta} by Li et al. makes iDFlakies evolution-aware and analyzes only tests that are affected by code changes to detect newly-introduced OD tests. As the best resulting method by Li et al.'s Intra-Class, we compared our work with their method's behavior. Gyori et al. proposed PolDet\cite{gyori_poldet} to detect tests that modify shared heap-state and do not reset it after
execution, meaning they are potential polluters. We used this approach in our work. In Python, Gruber et al. introduced Flapy\cite{flapy_gruber_icsec} to mine flaky tests in a given or automatically sampled set of Python projects by re-running their test suites.
For fixing OD tests, Shi et al. proposed iFixFlakies\cite{ishi_FixFlakies_fse} to automatically generate a patch for a known OD test. They used the tests in the test suite to find potential cleaners for an OD test and based on this generated a patch method. They found cleaner for 58 OD tests in 110 cases. It indicates that often there will be cleaner code in the test classes. From here we can explore whether the problem also lies in the test class or not. Our study finds out it can be as frequent as 23 out of 27 projects having all of its OD reasoned in the test classes. Li et al. later improved upon iFixFlakies in ODRepair\cite{li_odrepair_icse} to fix the OD tests that do not have a cleaner in test classes. They used the shared static fields found in the code under test to generate patches. Wang et al. developed iPFlakies\cite{wang_ipflakies_icsec} for Python projects to automatically repair Python OD tests.
Although having various studies in detecting and fixing OD, no prior work has been done on minimizing the required re-runs of test methods in detecting OD tests. In a recent study by Baz et al.\cite{baz_latest_ase}, it is found that execution time can vary based on the order of the suite. They reordered tests to search for the fastest test order by focusing on relative positioning between test classes in the test order. Li et al.\cite{li_latest_ase}, in another recent study, propose an approach for reducing test runtime by transforming test fixtures. They targeted test fixtures that run before/after individual tests defined in a test class to instead run once at the beginning and/or at the end of the test class, before/after all tests in the test class run. Parry et al. proposed Cannier\cite{parry_cannier_se}, an approach for reducing the time cost of re-running-based flaky test detection techniques by combining them with machine learning models.

Although these methods work to shorten the order runtime to detect OD tests fast, none works to prioritize the contributing tests. We focus on prioritizing tests to run to detect possible OD tests. And reducing tests will shorten the runtime also.

\section{Discussion and Future Works}
The result of our method shows a potential to be a vital part of test case prioritization. However, some questions may arise about the method's reliability.

We have used the only test classes to prioritize test cases. Our analysis of the projects implies that developers are usually less concerned about the shared memory of test classes rather than the source code. For this reason, static variables from the test class are often overlooked. We addressed this issue in our work, and it reflects the promise of prioritizing the most suspected ones.

Another issue can be raised on the reduction of test methods, reducing the test suit hampers the testing goal or not, according to the code coverage issue. Here, reducing test methods does not hamper the goal of testing. In terms of the objective, effectively detecting OD tests, code coverage is not an issue here. We can run the complete coverage testing with all the test cases whenever necessary. According to the objective of our study, removing test methods from orders is beneficial rather than problematic.

Prioritizing tests in OD detection can be an expansive area to work on. We have used the static fields' behavior of the test classes. Future works can look for more information from the test class and explore source code. We are currently working on analyzing the source code behavior and exploring the potential of it to contribute to test prioritization in order-dependent flaky test detection. We leave the data and code of this work open to all for further improvement\cite{hasnain_2024_14064478}.

\section{Conclusion}
In conclusion, flaky tests, especially those that are order-dependent (OD), reduce the reliability of software testing by generating unpredictable results. This inconsistency can mislead developers, causing unnecessary debugging efforts, and, in some cases, even hiding actual bugs. Although previous methods for identifying OD tests, such as random test ordering or systematic test pairing, have been useful, they are often costly in terms of time and computing resources due to the need for repeated test runs. Our approach addresses this problem by prioritizing tests that are most likely to be order-dependent, focusing on shared states in memory that commonly lead to flaky behavior. This method reduces the number of test re-runs required, achieving a 72.19\% re-run reduction on average across 27 projects. By streamlining the detection of OD tests, our approach improves the efficiency of testing and helps create more stable, reliable software.

\section{Acknowledgement}
This study is supported by the fellowship from the ICT Division, Government of Bangladesh – No.: 56.00.0000.052.33.0003.24-112, dated 03.06.2025.
\clearpage
\bibliographystyle{IEEEtran}
\bibliography{camera}

\end{document}